# On Event Reduction in Localization of DES Supervisory Control

Vahid Saeidi[1], Ali A. Afzalian[2] and Davood Gharavian[3]
Department of Electrical Eng., Abbaspour School of Engineering,
Shahid Beheshti University, Tehran, Iran
[1] v_saeidi@sbu.ac.ir, [2] Afzalian@sbu.ac.ir, [3] d_gharavian@sbu.ac.ir

**Abstract**

Supervisor localization procedure can be used to construct local controllers corresponding to each component agent in discrete-event systems. This procedure is based on state reduction of a monolithic supervisor with respect to each set of controllable events corresponding to each component agent. State reduction from the reduced supervisor to each local controller is an important criterion. In this paper, we deal with event reduction. It is proved that the number of events in each local controller is less than the event cardinality in the reduced supervisor, provided that each local controller has less number of states comparing to the reduced supervisor. It shows that each local controller needs less number of events to make consistent decisions. Having this property, we can evaluate the localizability of a supervisor by event reduction criteria instead of state reduction criteria. State reduction facilitates the implementation of local controllers on industrial systems. Whereas, event reduction reduces communication traffic between each pair of local controllers.

## 1. Introduction

The state size of the monolithic supervisory controller increases with state sizes of the plant and specification which increases the computational complexity and may lead to state explosion [1]. Thus, the application of this theory is restricted [2, 3]. Although, modular [4-7] and incremental [8, 9] approaches try to overcome the complexity of the supervisor synthesis, other approaches tend to reduce the supervisor for simple implementation [10]. Supervisor localization procedure, introduced in [11], is a method to distribute the supervisory control of discrete-event system. This procedure achieves two goals: i) it preserves the optimality and non-blocking of the monolithic supervisor, and ii) it dramatically simplifies each local controller based on the state size criteria. Namely, a supervisor is localizable if the state size of each local controller is less than the state size of the reduced state supervisor. This procedure is carried out using control information relevant to the target agent. As the result, each agent obtains its own local controller. Some applications of supervisor localization can be found in [12].

Supervisor localization procedure can be employed to construct a distributed supervisory control for a monolithic specification associated with component agents. To the best of the author's knowledge, no work has been carried out on event reduction properties of this procedure. On the other hand, decomposition of a supervisor [13] is an alternative method to reduce the number of events, considered in decision making by each decentralized supervisor in large scale discrete-event systems (DES). This method constructs a distributed supervisory control with restricted control authority and restricted observation scope.

In this paper, we investigate the event reduction in the localization of a monolithic supervisor. The main common concepts in the supervisor reduction and supervisor localization procedures are control consistency of states, control cover, and normality of an induced generator which is control equivalent to the supervisor w.r.t. the plant. Moreover, a constructed induced generator is deterministic automata.

We prove that, if a supervisor is localizable, (i.e. the state size of each local controller corresponding to each component of the plant is less than the state cardinality of the reduced supervisor), then at least one controllable event corresponding to other components appears as self loop transition at all states of the corresponding local controller (in the sense of normality [14, 15]). It is an important result. However, event reduction in each local controller does not guarantee decomposability of the supervisor. Moreover, each local controller needs fewer events for consistent decision making, comparing to the reduced state supervisor. It reduces the communication traffic between local controllers in a distributed supervisory control. Examples 4.2 and 4.3, show that decomposition of a supervisor is stronger than event reduction in each local controller of a distributed supervisory control.

The rest of the paper is organized as follows: In Section 2, the necessary preliminaries are reviewed. In this section, we review the formulation of supervisor reduction and supervisor localization procedures, by which one can prove required propositions. In Section 3, we prove that, state reduction leads to event reduction in each local controller. In Section 4, we compare state reduction and event reduction by some examples, and show that event reduction is more practical criteria for localizability of a supervisor. Finally, concluding remarks and future work are given in Section 5.

## 2. Preliminaries

A discrete-event system is presented by an automaton $\mathbf{G} = (Q, \Sigma, \delta, q_0, Q_m)$, where $Q$ is a finite set of states, with $q_0 \in Q$ as the initial state and $Q_m \subseteq Q$ being the marker states; $\Sigma$ is a finite set of events ($\sigma$) which is partitioned as a set of controllable events $\Sigma_c$ and a set of uncontrollable events $\Sigma_u$, where $\Sigma = \Sigma_c \cup \Sigma_u$. $\delta$ is a transition mapping $\delta: Q \times \Sigma \to Q$, $\delta(q, \sigma) = q'$ gives the next state $q'$ is reached from $q$ by the occurrence of $\sigma$. $\mathbf{G}$ is discrete-event model of the plant. In this context $\delta(q_0, s)!$ means that $\delta$ is defined for $s$ at $q_0$. $L(\mathbf{G}) := \{s \in \Sigma^* | \delta(q_0, s)!\}$ is the closed behavior of $\mathbf{G}$ and $L_m(\mathbf{G}) := \{s \in L(\mathbf{G}) | \delta(q_0, s) \in Q_m\}$ is the marked behavior of $\mathbf{G}$ [17, 18]. In this paper, we assume that $\mathbf{G}$ consists of component agents $\mathbf{G}^k$ defined on pairwise disjoint events sets $\Sigma^k$ ($k \in \mathcal{K}$, $\mathcal{K}$ an index set), i.e. $\Sigma = \cup\{\Sigma^k | k \in \mathcal{K}\}$. Let $L_k := L(\mathbf{G}^k)$ and $L_{m,k} := L_m(\mathbf{G}^k)$, the closed and marked behaviors of $\mathbf{G}$ are $L(\mathbf{G}) = \| \{L_k | k \in \mathcal{K}\}$ and $L_m(\mathbf{G}) = \| \{L_{m,k} | k \in \mathcal{K}\}$. We assume that for every $k \in \mathcal{K}$, $\overline{L}_{m,k} = L_k$ is true. Then $\mathbf{G}$ is necessarily nonblocking (i.e. $\overline{L}_m(\mathbf{G}) = L(\mathbf{G})$).

A set of all control patterns is denoted with $\Gamma = \{\gamma \in Pwr(\Sigma) | \gamma \supseteq \Sigma_u\}$. A supervisor of a plant $\mathbf{G}$, is a map $V: L(\mathbf{G}) \to \Gamma$ where $V(s)$ represents the set of enabled events after the occurrence of the string $s \in L(\mathbf{G})$. A pairs $(\mathbf{G}, V)$ is written by $V/\mathbf{G}$ and called $\mathbf{G}$ is under supervision by $V$. A behavioral constraint on $\mathbf{G}$ is given by specification language $E \subseteq \Sigma^*$. Let $K \subseteq L_m(\mathbf{G}) \cap E$ be the supremal controllable sublanguage of $E$ w.r.t. $L(\mathbf{G})$ and $\Sigma_u$, i.e. $K = supC(L_m(\mathbf{G}) \cap E)$. If $K \neq \emptyset$, $\mathbf{SUP} = (X, \Sigma, \xi, x_0, X_m)$ is recognizer of $K$. If $\mathbf{G}$ and $\mathbf{E}$ are finite-state DES, then $K$ is regular language. Write $|.|$ for the state size of DES. Then $|\mathbf{SUP}| \leq |\mathbf{G}||\mathbf{E}|$. In applications, engineers want to employ the reduced supervisor $\mathbf{RSUP}$, which has a fewer number of states ( i.e. $|\mathbf{RSUP}| \ll |\mathbf{SUP}|$) and is control equivalent to $\mathbf{SUP}$ w.r.t. $\mathbf{G}$ [10], i.e.

$$L_m(\mathbf{G}) \cap L_m(\mathbf{RSUP}) = L_m(\mathbf{SUP}) \tag{1}$$
$$L(\mathbf{G}) \cap L(\mathbf{RSUP}) = L(\mathbf{SUP}) \tag{2}$$

A generator $\mathbf{LOC}^k$ over $\Sigma$, is a local controller for agent $\mathbf{G}^k$, if $\mathbf{LOC}^k$ can disable only events in $\Sigma_c^k$. Precisely, for all $s \in \Sigma^*$ and $\sigma \in \Sigma$, there holds

$$s\sigma \in L(\mathbf{G}) \ \& \ s \in L(\mathbf{LOC}^k) \ \& \ s\sigma \notin L(\mathbf{LOC}^k) \Rightarrow \sigma \in \Sigma_c^k$$

The observation scope of $\mathbf{LOC}^k$ is not limited to $\Sigma^k$. But, the control authority of a local controller is strictly local [11].

A set of local controllers $\mathbf{LOC} = \{\mathbf{LOC}^k | k \in \mathcal{K}\}$ is constructed, each one for an agent, with $L(\mathbf{LOC}) = \cap \{L(\mathbf{LOC}^k) | k \in \mathcal{K}\}$ and $L_m(\mathbf{LOC}) = \cap \{L_m(\mathbf{LOC}^k) | k \in \mathcal{K}\}$ such that the following relationships hold,

$$L_m(\mathbf{G}) \cap L_m(\mathbf{LOC}) = L_m(\mathbf{SUP}) \tag{3}$$
$$L(\mathbf{G}) \cap L(\mathbf{LOC}) = L(\mathbf{SUP}) \tag{4}$$

We say that, **LOC** is control equivalent to **SUP** w.r.t. **G**, if (3) and (4) are satisfied. This formulation is based on state reduction of a monolithic supervisor with respect to disabled controllable events of each component agent.

The natural projection is a mapping $P: \Sigma^* \to \Sigma_0^*$ where (1) $P(\epsilon) := \epsilon$, (2) for $s \in \Sigma^*$, $\sigma \in \Sigma$, $P(s\sigma) := P(s)P(\sigma)$, and (3) $P(\sigma) := \sigma$ if $\sigma \in \Sigma_0$ and $P(\sigma) := \epsilon$ if $\sigma \notin \Sigma_0$. The effect of an arbitrary natural projection $P$ on a string $s \in \Sigma^*$ is to erase the events in $s$, that do not belong to observable events set, $\Sigma_0$. The natural projection $P$ can be extended and denoted by $P: Pwr(\Sigma^*) \to Pwr(\Sigma_0^*)$. For any $L \subseteq \Sigma^*$, $P(L) := \{P(s)|s \in L\}$. The inverse image function of $P$ is denoted by $P^{-1}: Pwr(\Sigma_0^*) \to Pwr(\Sigma^*)$ for any $L \subseteq \Sigma_0^*$, $P^{-1}(L) := \{s \in \Sigma^* | P(s) \in L\}$. The synchronous product of languages $L_1 \subseteq \Sigma_1^*$ and $L_2 \subseteq \Sigma_2^*$ is defined by $L_1 \parallel L_2 = P_1^{-1}(L_1) \cap P_2^{-1}(L_2) \subseteq \Sigma^*$, where $P_i: \Sigma^* \to \Sigma_i^*$, $i = 1,2$ for the union $\Sigma = \Sigma_1 \cup \Sigma_2$ [7].

It was defined in [16], that $K$ is relative observable w.r.t. $\bar{C}, \mathbf{G}$ and $P$ (or simply $\bar{C}$-observable) for $K \subseteq C \subseteq L_m(\mathbf{G})$, where $\bar{K}$ and $\bar{C}$ are prefix closed languages, if for every pair of strings $s, s' \in \Sigma^*$ such that $P(s) = P(s')$, the following two conditions hold:

- $(\forall \sigma \in \Sigma)\, s\sigma \in \bar{K}, s' \in \bar{C}, s'\sigma \in L(\mathbf{G}) \Rightarrow s'\sigma \in \bar{K}$
- $s \in K, s' \in \bar{C} \cap L_m(\mathbf{G}) \Rightarrow s' \in K$

In the special case, if $C = K$, then the relative observability property is tighten to the observability property. An observation property called normality was defined in [14], that is stronger than the relative observability. $K$ is said to be normal w.r.t. $(L(\mathbf{G}), P)$, if $P^{-1}P(\bar{K}) \cap L(\mathbf{G}) = \bar{K}$, where $L(\mathbf{G})$ is a prefix closed language and $P$ is a natural projection.

### 2.1. Supervisor reduction procedure

A procedure was proposed in [10], to reduce the state size of a monolithic supervisor. This method constructs a generator which is control equivalent to the monolithic supervisor w.r.t. the plant. Let $\mathbf{SUP} = (X, \Sigma, \xi, x_0, X_m)$ and define $E: X \to Pwr(\Sigma)$ as $E(x) = \{\sigma \in \Sigma | \xi(x, \sigma)!\}$. $E(x)$ denotes the set of events enabled at state $x$. Next, define $D: X \to Pwr(\Sigma)$ as $D(x) = \{\sigma \in \Sigma | \neg \xi(x, \sigma)! \,\&\, (\exists s \in \Sigma^*)[\xi(x_0, s) = x \,\&\, \delta(q_0, s\sigma)!]\}$. $D(x)$ is the set of events which are disabled at state $x$. Define $M: X \to \{1,0\}$ according to $M(x) = 1$ iff $x \in X_m$, namely the flag of $M$ determines if a state is marked in **SUP**. Also, define $T: X \to \{1,0\}$ according to $T(x) = 1$ iff $(\exists s \in \Sigma^*)\xi(x_0, s) = x \,\&\, \delta(q_0, s) \in Q_m$, namely the flag of $T$ determines if some corresponding state is marked in **G**. Let $\mathcal{R} \subseteq X \times$

$X$ be the binary relation such that for $x, x' \in X$, $(x, x') \in \mathcal{R}$. $x$ and $x'$ are called control consistent, if

$$E(x) \cap D(x') = E(x') \cap D(x) = \emptyset \tag{5}$$
$$T(x) = T(x') \Rightarrow M(x) = M(x') \tag{6}$$

Informally, a pair of $(x, x')$ is in $\mathcal{R}$ if, by (3), there is no event enabled at $x$ but disabled at $x'$, and by (4), $(x, x')$ are both marked (unmarked) in **SUP**, provided that they are both marked (unmarked) in **G**. While $\mathcal{R}$ is reflexive and symmetric, it need not be transitive, consequently it is not an equivalence relation. This fact underlies the next definition. A cover $\mathcal{C} = \{X_i \subseteq X | i \in I\}$ of $X$ is called a control cover on **SUP** if [10],

$$(\forall i \in I) X_i \neq \emptyset \wedge (\forall x, x' \in X_i)(x, x') \in \mathcal{R} \tag{7}$$
$$(\forall i \in I)(\forall \sigma \in \Sigma)(\exists j \in I)\big[(\forall x \in X_i)\xi(x, \sigma)! \Rightarrow \xi(x, \sigma) \in X_j\big], \tag{8}$$

Where, $I$ is an index set.

A control cover $\mathcal{C}$ lumps states of **SUP** into cells $X_i$ ($i \in I$) if they are control consistent. A control cover $\mathcal{C}$ is control congruence if $X_i$ are pairwise disjoint. According to (5), each cell of $\mathcal{C}$ is nonempty and each pair of states in one cell should be consistent. According to (6), all states that can be reached from any states in $X_i$ by one step transition $\sigma$ is covered by some $X_j$.

Given control cover $\mathcal{C} = \{X_i \subseteq X | i \in I\}$ on **SUP**, an induced supervisor is constructed as $\mathbf{J} = (I, \Sigma, \kappa, i_0, I_m)$ where $i_0 =$ some $i \in I$ with $x_0 \in X_i$, $I_m = \{i \in I | X_i \cap X_m \neq \emptyset\}$ and $\kappa: I \times \Sigma \to I$ with $\kappa(i, \sigma) = j$ provided, for such choice of $j \in I$,

$$(\exists x \in X_i)\xi(x, \sigma) \in X_j \ \& \ (\forall x' \in X_i)[\xi(x', \sigma)! \Rightarrow \xi(x', \sigma) \in X_j] \tag{9}$$

Overlapping of some states results that $i_0$ and $\kappa$ may not be uniquely determined, thus $\mathbf{J}$ may not be unique. If $\mathcal{C}$ is control congruence, then $\mathbf{J}$ is uniquely determined by $\mathcal{C}$. Generally, $\mathbf{J}$ is control equivalent to **SUP** w.r.t **G**.

A DES **RSUP** $= (Z, \Sigma, \zeta, z_0, Z_m)$ is normal w.r.t **SUP** if,

$(i)(\forall z \in Z)(\exists s \in L(\mathbf{SUP}))\zeta(z_0, s) = z$
$(ii)(\forall z \in Z)(\forall \sigma \in \Sigma)[\zeta(z, \sigma)! \Rightarrow (\exists s \in L(\mathbf{SUP}))[s\sigma \in L(\mathbf{SUP}) \ \& \ \zeta(z_0, s) = z]]$ (10)
$(iii)(\forall z \in Z_m)(\exists s \in L_m(\mathbf{SUP}))\zeta(z_0, s) = z$

Given two generators **RSUP** $= (Z, \Sigma, \zeta, z_0, Z_m)$ and $\mathbf{J} = (I, \Sigma, \kappa, i_0, I_m)$ are DES-isomorphic with isomorphism $\theta$ if there exists a map $\theta: Z \to I$ such that

(i) $\theta: Z \to I$ is a bijection

(ii) $\theta(z_0) = i_0$ and $\theta(Z_m) = I_m$

(iii) $(\forall z \in Z)(\forall \sigma \in \Sigma)\zeta(z,\sigma)! \Rightarrow [\kappa(\theta(z),\sigma)! \,\&\, \kappa(\theta(z),\sigma) = \theta(\zeta(z,\sigma))]$

(iv) $(\exists i \in I)(\forall \sigma \in \Sigma)\kappa(i,\sigma)! \Rightarrow [(\exists z \in Z)\zeta(z,\sigma)! \,\&\, \theta(z) = i]$

(11)

It was proved in [10], if **SUP** is supremal supervisor for **G** and **RSUP** is any normal supervisor w.r.t **SUP** such that it is control equivalent to **SUP** w.r.t **G**, then there exists a control cover $\mathcal{C}$ on **SUP** for which some induced supervisor **J** is DES-isomorphic to **RSUP**.

### 2.2. Supervisor localization procedure

A procedure was proposed in [11], to construct a set of local controllers which their synchronization with the plant is equivalent to the monolithic supervisor. This method follows from $\Sigma = \bigcup\{\Sigma^k | k \in \mathcal{K}\}$ that the set $\{\Sigma_c^k \subseteq \Sigma_c | k \in \mathcal{K}\}$ make a partition on $\Sigma_c$. This method employs control consistency and control cover from supervisor reduction procedure, proposed in [10].

Define $E: X \to Pwr(\Sigma)$ as $E(x) = \{\sigma \in \Sigma | \xi(x,\sigma)!\}$. Next, define $D^k: X \to Pwr(\Sigma_c^k)$ as $D^k(x) = \{\sigma \in \Sigma_c^k | \neg \xi(x,\sigma)! \,\&\, (\exists s \in \Sigma^*)[\xi(x_0,s) = x \,\&\, \delta(q_0,s\sigma)!]\}$. The flag of $M$ and $T$ are defined same as the ones defined in the supervisor reduction procedure. Let $\mathcal{R}^k \subseteq X \times X$ be the binary relation such that for $x, x' \in X$, $(x, x') \in \mathcal{R}^k$. $x$ and $x'$ are called control consistent w.r.t. $\Sigma_c^k$, if

$$E(x) \cap D^k(x') = E(x') \cap D^k(x) = \emptyset \qquad (12)$$

$$T(x) = T(x') \Rightarrow M(x) = M(x') \qquad (13)$$

A cover $\mathcal{C}^k = \{X_{i^k}^k \subseteq X | i^k \in I^k\}$ is a control cover on $X$ w.r.t. $\Sigma_c^k$ if,

$$(\forall i^k \in I^k)\, X_{i^k}^k \neq \emptyset \land (\forall x, x' \in X_{i^k}^k)(x, x') \in \mathcal{R}^k \qquad (14)$$

$$(\forall i^k \in I^k)(\forall \sigma \in \Sigma)(\exists j^k \in I^k)\left[(\forall x \in X_{i^k}^k)\xi(x,\sigma)! \Rightarrow \xi(x,\sigma) \in X_{j^k}^k\right], (15)$$

Where, $I^k$ is some index set [11].

Given a control cover $\mathcal{C}^k$ on $X$, based only on the control information of $\Sigma_c^k$, an induced generator $\mathbf{J}^k = (I^k, \Sigma, \kappa^k, i_0^k, I_m^k)$ is obtained by the following construction

(i) $i_0^k \in I^k$ such that $x_0 \in X_{i_0^k}^k$

(ii) $I_m^k = \{i^k \in I^k | X_{i^k}^k \cap X_m \neq \emptyset\}$

(iii) $\kappa^k: I^k \times \Sigma \to I^k$ with $\kappa^k(i^k, \sigma) = j^k$, if

$(\exists x \in X_{i^k}^k)\xi(x,\sigma) \in X_{j^k}^k \,\&\, (\forall x' \in X_{i^k}^k)[\xi(x',\sigma)! \Rightarrow \xi(x',\sigma) \in X_{j^k}^k]$

(16)

Overlapping of some states results that $i_0^k$ and $\kappa^k$ may not be uniquely determined, and $\mathbf{J}^k$ may not be unique. If $\mathcal{C}^k$ is control congruence, then $\mathbf{J}^k$ is uniquely determined by $\mathcal{C}^k$. We can obtain a set of induced generators $\mathbf{J} = \{\mathbf{J}^k | k \in \mathcal{K}\}$. Let $L(\mathbf{J}) := \cap \{L(\mathbf{J}^k) | k \in \mathcal{K}\}$ and $L_m(\mathbf{J}) := \cap \{L_m(\mathbf{J}^k) | k \in \mathcal{K}\}$. $\mathbf{J}$ is a solution to the distributed supervisory control problem.

A DES $\mathbf{LOC} = (Z_L, \Sigma, \zeta_L, z_{L,0}, Z_{L,m})$ is normal w.r.t $\mathbf{SUP}$ if,

$$(i)(\forall z \in Z_L)(\exists s \in L(\mathbf{SUP}))\zeta(z_{L,0}, s) = z$$
$$(ii)(\forall z \in Z_L)(\forall \sigma \in \Sigma)[\zeta(z, \sigma)! \Rightarrow (\exists s \in L(\mathbf{SUP}))[s\sigma \in L(\mathbf{SUP}) \& \zeta(z_{L,0}, s) = z]] \quad (17)$$
$$(iii)(\forall z \in Z_{L,m})(\exists s \in L_m(\mathbf{SUP}))\zeta(z_{L,0}, s) = z$$

Given two generators $\mathbf{LOC} = (Z_L, \Sigma, \zeta_L, z_{L,0}, Z_{L,m})$ and $\mathbf{J} = (I, \Sigma, \kappa, i_0, I_m)$ are DES-isomorphic with isomorphism $\theta$, if there exists a map $\theta: Z_L \to I$ such that

$$\begin{aligned}
&(i) \ \theta: Z_L \to I \text{ is a bijection} \\
&(ii) \ \theta(z_{L,0}) = i_0 \text{ and } \theta(Z_{L,m}) = I_m \\
&(iii) \ (\forall z \in Z_L)(\forall \sigma \in \Sigma)\zeta_L(z, \sigma)! \Rightarrow [\kappa(\theta(z), \sigma)! \& \kappa(\theta(z), \sigma) = \theta(\zeta_L(z, \sigma))] \\
&(iv)(\exists i \in I)(\forall \sigma \in \Sigma)\kappa(i, \sigma)! \Rightarrow [(\exists z \in Z_L)\zeta_L(z, \sigma)! \& \theta(z) = i]
\end{aligned} \quad (18)$$

Under normality and DES-isomorphism, let $\mathbf{LOC}$ be a set of normal generators that is control equivalent to $\mathbf{SUP}$ w.r.t. $\mathbf{G}$. Then, there exists a set of control covers $\mathcal{C} := \{\mathcal{C}^k | k \in \mathcal{K}\}$ on $X$ with a corresponding set of induced generators $\mathbf{J} := \{\mathbf{J}^k | k \in \mathcal{K}\}$ such that, for every $k \in \mathcal{K}$, $\mathbf{J}^k$ and $\mathbf{LOC}^k$ are DES-isomorphic.

### 3. Event reduction in local controllers

The aim of supervisor localization procedure is to reduce the state size of a supervisor corresponding to each component of the plant. Thus, when a pair of states are control consistent in $\mathcal{C}$, they are also control consistent in $\mathcal{C}^k, \forall k \in \mathcal{K}$. On the other hand, if the state size of a local controller is less than the state cardinality of the reduced supervisor, then a pair of states which are not control consistent in $\mathcal{C}$ may be control consistent in $\mathcal{C}^k$, where $k \in \mathcal{K}$. In the following proposition, we prove that, if a local controller has fewer states comparing to the reduced supervisor, then there exists at least one self-looped event, which occurs at one state of the local controller.

*Proposition 1:* Let $\mathbf{G}$ be a non-blocking plant, consists of components $\mathbf{G}^k, k \in \mathcal{K}$, $\mathbf{SUP}$ be a supervisor of $\mathbf{G}$, $\mathbf{RSUP}$ be the reduced supervisor, and $\mathbf{LOC}^k$ be a local controller corresponding to $\mathbf{G}^k$. If $\mathbf{LOC}^k$ has less number of states than $\mathbf{RSUP}$, then $\exists \sigma \in \Sigma_c^j, j \neq k$ such that $\sigma$ is self-looped at one state of $\mathbf{LOC}^k$.

*Proof:* Assume that $x_i \in X$, such that $(\exists s_i \in \Sigma^*), x_i = \xi(x_0, s_i)$.

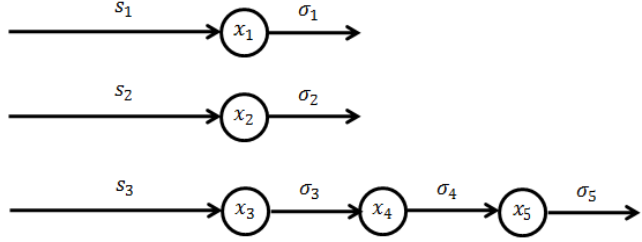

Fig. 1. A set of states and strings in **SUP**

From the typical setup, shown in Fig. 1, we substitute lumped control consistent states $x_1, x_2$ and $x_1, x_3$, by $A$ and $B$, respectively in **RSUP** (Fig. 2).

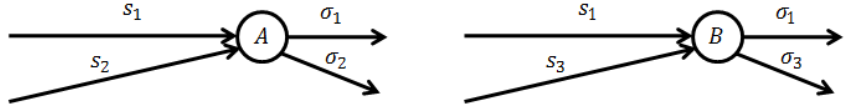

Fig. 2. A set of states and strings in **RSUP**

Also, we substitute lumped control consistent states $x_1, x_2, x_3$ by $M$ in **LOC**$^k$ (Fig. 3).

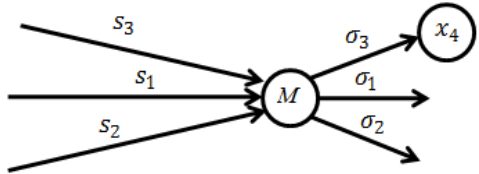

Fig. 3. A set of states and strings in **LOC**$^k$

We prove that $\sigma_3$ is self-looped at state $M$.

In Fig. 1, assume that $x_3$ is not control consistent with $x_2$. Since (6) and (13) are the same, we employ only (5) and (12) for proving the claim. Thus, we can write,

$$E(x_1) \cap D(x_3) = E(x_3) \cap D(x_1) = \emptyset \Rightarrow \sigma_3 \notin D(x_1), \text{ and } \sigma_1 \notin D(x_3) \quad (19)$$
$$E(x_2) \cap D(x_3) \neq \emptyset \text{ or } E(x_3) \cap D(x_2) \neq \emptyset \Rightarrow \sigma_3 \in D(x_2), \text{ or } \sigma_2 \in D(x_3) \quad (20)$$

From (19), we can write,

If $\sigma_3 \in D(x_2)$, then $\sigma_3 \notin E(x_1)$. Because $x_1$ and $x_2$ are control consistent. Thus, $s_1\sigma_3 \notin L(\mathbf{G})$.

From (20), we can write,

If $\sigma_2 \in D(x_3)$, then $\sigma_2 \notin E(x_1)$. Because $x_1$ and $x_3$ are control consistent. Thus, $s_1\sigma_2 \notin L(\mathbf{G})$.

From (19) and (20), we can write,

If $\sigma_3 \in D(x_2)$, and $\sigma_2 \in D(x_3)$, then $\sigma_2, \sigma_3 \notin E(x_1)$. Thus, $s_1\sigma_2, s_1\sigma_3 \notin L(\mathbf{G})$.

Let $\sigma_3 \in D(x_2)$. It means that $\sigma_3$ is a controllable event. But, $\sigma_3 \notin D^k(x_2)$. It means that $s_2\sigma_3 \in L(\mathbf{G})$, the monolithic supervisor disables $\sigma_3$ at $x_2$, and the local controller is not authorized to disable $\sigma_3$ at $x_2$. Hence, $\sigma_3$ is not an enabled transition by the monolithic supervisor at $x_2$, and it must be self-looped at state $M$ in $\mathbf{LOC}^k$ (Fig. 4).

From (17), a set of local controllers, constructed by supervisor localization procedure, are normal w.r.t. **SUP**. It means that, each string belongs to a local controller and the plant must be enabled in the monolithic supervisor. Thus, $\sigma_3$ cannot be self-looped at state $M$, while being a transition to $x_4$. Therefore, $\sigma_3$ is self-looped at state $M$, only.

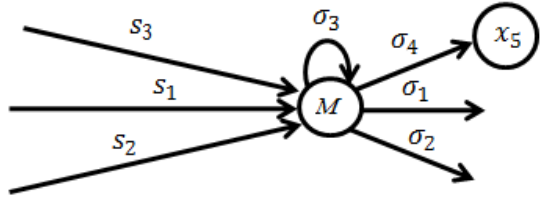

Fig. 4. A set of states and strings in $\mathbf{LOC}^k$, where $\sigma_3$ is self-looped at $M$

□

*Corollary 1:* In Proposition 1, since $\sigma_3$ is disabled at some states of the supervisor, its observation affects the behavior of the supervisor. It means that, if $\sigma_3$ is not observed, then the supervisor make inconsistent decisions (in the sense of normality).

In order to prove the main claim of the paper, we show that $\sigma_3$ appears as a self loop transition at all states of $\mathbf{LOC}^k$. Since $\mathbf{LOC}^k$ is not authorized to disable the controllable event $\sigma_3$, it is sufficient to prove that $\sigma_3$ appears as a self loop transition at states, where it is enabled. In Lemma 1, we prove the claim in a special case.

*Lemma 1:* Let $\mathbf{G}$ be a non-blocking plant, consists of components $\mathbf{G}^k, k \in \mathcal{K}$, and $\mathbf{LOC}^k$ be a local controller corresponding to each $\mathbf{G}^k$. Let $\sigma_0 \in \Sigma_c^j, j \neq k$, and one string belongs to $L(\mathbf{LOC}^k)$ be as shown in Fig. 5. If $[(\exists s \in \Sigma^*), \delta(q_0, s\sigma_0)!, \neg\delta(q_0, s\sigma_1)!, \exists i, z_i = \zeta_L(z_{L,0}, s) \Rightarrow z_i = \zeta_L(z_i, \sigma_0)]$, then $[\forall i, \zeta_L(z_i, \sigma_0)! \Rightarrow z_i = \zeta_L(z_i, \sigma_0)]$.

*Proof:* Assume the set of states and strings, are shown in Fig. 5. Define $P: \Sigma^* \to \Sigma_0^*$, $\Sigma_0 = \Sigma - \{\sigma_0\}$. We should prove that each pair of states where $\sigma_0$ occurs between them, can be considered one state. We know that $P(s\sigma_0\sigma_1s'\sigma_0) = P(s\sigma_1s')$. Since a language, which is constructed by local controllers in the plant are same as the language of the monolithic supervisor (see (3) and (4)), we can extend the observability property of the monolithic supervisor to each local controller. Thus, from Fig. 5, we can write,

$\forall \sigma \in \Sigma, s\sigma_0\sigma_1s'\sigma_0\sigma \in L(\mathbf{LOC}^k) \cap L(\mathbf{G}), s\sigma_1s' \in L(\mathbf{LOC}^k), s\sigma_1s'\sigma \in L(\mathbf{G}) \Rightarrow s\sigma_1s'\sigma \in L(\mathbf{LOC}^k)$ (21)

The string, which occurs in both the local controller and the plant is shown in the first term of the antecedent in (21). Since $s\sigma_1 \notin L(\mathbf{G})$, then $s\sigma_1 s' \notin L(\mathbf{G})$. Thus, (21) is true. Similarly, we can write,

$$s\sigma_0\sigma_1 s'\sigma_0 \in L_m(\mathbf{LOC}^k) \cap L_m(\mathbf{G}), s\sigma_1 s' \in L(\mathbf{LOC}^k) \cap L_m(\mathbf{G}) \Rightarrow s\sigma_1 s' \in L_m(\mathbf{LOC}^k) \tag{22}$$

Since $s\sigma_1 s' \notin L(\mathbf{G})$, then $s\sigma_1 s' \notin L_m(\mathbf{G})$. Thus, (22) is true. Therefore, $z_n$ and $z_{n+1}$ can be considered one state, where $\sigma_0$ is a self-looped transition.

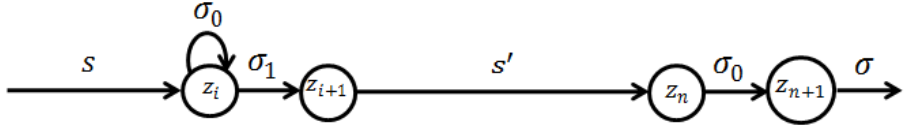

Fig. 5. A set of states and strings in $\mathbf{LOC}^k$, where $\sigma_0$ is self-looped at $z_i$

However, the supervisor reduction and the supervisor localization procedures consider the plant cyclic (the task of the plant is assumed cyclic), even if it is not cyclic (see example 4.2), we assume $\mathbf{G}$ is a cyclic plant. Thus, $\forall i, \zeta_L(z_i, \sigma_0)!$ we use the above argument and conclude that $z_i = \zeta_L(z_i, \sigma_0)$. Since $\mathbf{LOC}^k$ is not authorized to disable $\sigma_0$, it can be considered a self loop transition at other states, where $\sigma_0$ is not defined in $\mathbf{G}$, correspondingly, i.e.

$$(\exists s \in \Sigma^*)[\delta(q_0, s) = q \,\&\, z_i = \zeta_L(z_{L,0}, s) \,\&\, \neg\delta(q, \sigma_0)!] \Rightarrow z_i = \zeta_L(z_i, \sigma_0)$$

Therefore, $\sigma_0$ is self-looped at all states in $\mathbf{LOC}^k$. Informally, $L_m(\mathbf{LOC}^k)$ is normal.

□

Now, we relax the assumption $s\sigma_1 \notin L(\mathbf{G})$, and show that Lemma 1 still does hold.

*Theorem 1:* Let $\mathbf{G}$ be a non-blocking plant, consists of components $\mathbf{G}^k, k \in \mathcal{K}$, described by closed and marked languages $L(\mathbf{G}), L_m(\mathbf{G}) \subseteq \Sigma^*$ and $\mathbf{SUP} = (X, \Sigma, \xi, x_0, X_m)$ be the supervisor of $\mathbf{G}$. From Proposition 1, If $\sigma_3$ is self-looped at one state $M$ in $\mathbf{LOC}^k, k \in \mathcal{K}$, then $\sigma_3$ is self-looped at all states in $\mathbf{LOC}^k$.

*Proof:* Following the proof in Lemma 1, clearly, if $s_3\sigma_4 \notin L(\mathbf{G})$, then (23) and (24) are true, in Fig. 6.

$$s_3\sigma_3\sigma_4 s'\sigma_3\sigma \in L(\mathbf{LOC}^k) \cap L(\mathbf{G}), s_3\sigma_4 s' \in L(\mathbf{LOC}^k), s_3\sigma_4 s'\sigma \in L(\mathbf{G}) \Rightarrow s_3\sigma_4 s'\sigma \in L(\mathbf{LOC}^k) \tag{23}$$

$$s_3\sigma_3\sigma_4 s'\sigma_3 \in L_m(\mathbf{LOC}^k) \cap L_m(\mathbf{G}), s_3\sigma_4 s' \in L(\mathbf{LOC}^k) \cap L_m(\mathbf{G}) \Rightarrow s_3\sigma_4 s' \in L_m(\mathbf{LOC}^k) \tag{24}$$

Now, we want to prove that (23) and (24) are true, even if $s_3\sigma_4 \in L(\mathbf{G})$. From Figs. 4, 6, we know that $s_3\sigma_4 \in L(\mathbf{LOC}^k)$. Thus, $\sigma_4$ is enabled at state $M$ in $\mathbf{LOC}^k$. Since $\mathbf{LOC}^k$ is deterministic automata, $\sigma_4$ is a transition from $M$ to only one state in $\mathbf{LOC}^k$. There may be two cases related to $\sigma_4$: (a) $\sigma_4$ is a transition from $x_3$ to $x_5$, in $\mathbf{SUP}$ (Fig. 4), (b) $\sigma_4$ is a transition from $x_3$ to another state except $x_5$, in $\mathbf{SUP}$.

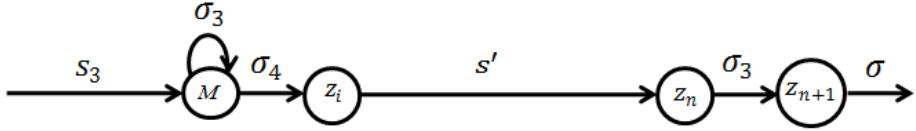

Fig. 6. A set of states and strings in $\mathbf{LOC}^k$, where $\sigma_3$ is self-looped at $M$

(a) $\underline{\sigma_4 \text{ is a transition from } x_3 \text{ to } x_5}$: If $\sigma_3$ is enabled at $x_4$, then $x_3, x_4$ are control consistent and $x_4$ and $M$ become one state in $\mathbf{LOC}^k$. In this case, we have a transition $\sigma_3$ from $M$ to another state (e.g. $x_6$), which is shown in Fig. 7(a). But, such transition makes $\mathbf{LOC}^k$ a non-deterministic automata. Thus, $x_6$ and $M$ must become one state (Fig. 7(b)). Obviously, $\sigma_3$ cannot be disabled at $x_6$, because $\mathbf{LOC}^k$ is not authorized to disable it.

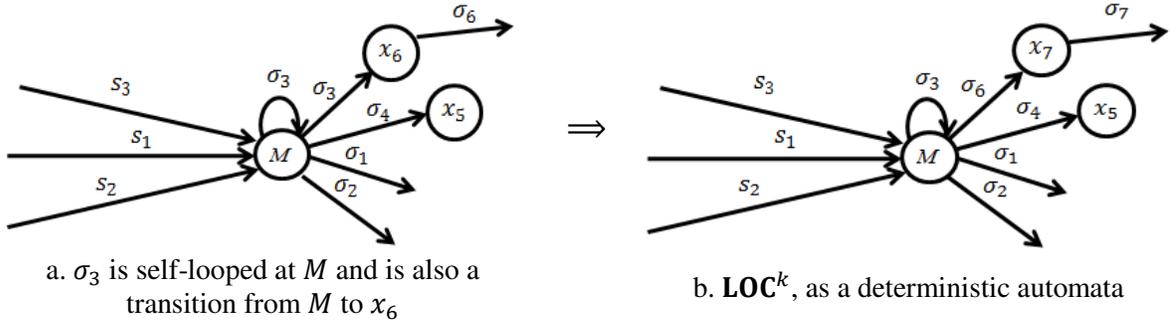

a. $\sigma_3$ is self-looped at $M$ and is also a transition from $M$ to $x_6$

b. $\mathbf{LOC}^k$, as a deterministic automata

Fig. 7. Transition $\sigma_3$ from $M$ to $x_6$, in $\mathbf{LOC}^k$

Also, $\sigma_3$ may be a transition from $M$ to $x_5$, which is shown in Fig. 8(a). In this case, $\mathbf{LOC}^k$ is also a non-deterministic automata. Thus, $x_5$ and $M$ must become one state (Fig. 8(b)), and $\sigma_4$ is also a self loop transition at $M$.

This argument can be continued for other states which may be control consistent with $M$, until either all subsequent transitions become self-looped at $M$ (in this case, $\mathbf{LOC}^k$ is reduced to one state with self loop transitions, and the claim is proved), or some enabled events can be found at a control consistent state with state $M$, such that they are not defined at corresponding state in $\mathbf{G}$. The latter was proved in Lemma 1. It means that, there is a natural projection $P: \Sigma^* \to \Sigma_0^*$, $\Sigma_0 = \Sigma_i \cup \{\sigma_3\}$ and $P(s_3\sigma_3\sigma_4 \dots \sigma_{n-1}) = P(s_3\sigma_4 \dots \sigma_{n-1})$, such that $\exists \sigma_n \in \Sigma, s_3\sigma_4 \dots \sigma_{n-1}\sigma_n \notin L(\mathbf{G})$ and

$s_3\sigma_3\sigma_4\ldots\sigma_{n-1}\sigma_n \in L(\mathbf{LOC}^k) \cap L(\mathbf{G}), s_3\sigma_4\ldots\sigma_{n-1} \in L(\mathbf{LOC}^k), s_3\sigma_4\ldots\sigma_{n-1}\sigma_n \in L(\mathbf{G})$
$\Rightarrow s_3\sigma_4\ldots\sigma_{n-1}\sigma_n \in L(\mathbf{LOC}^k)$

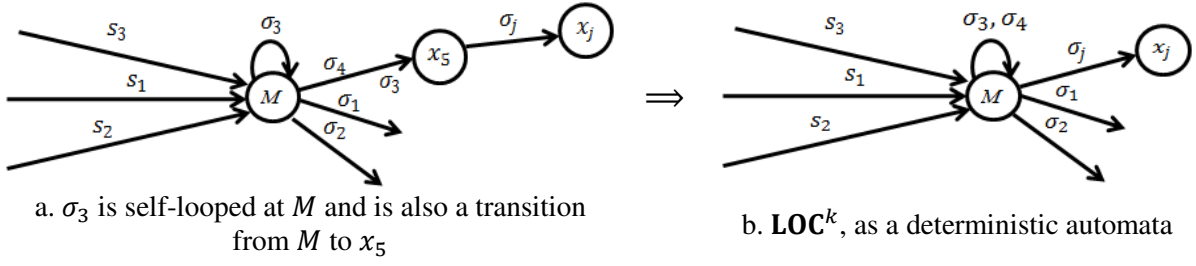

a. $\sigma_3$ is self-looped at $M$ and is also a transition from $M$ to $x_5$

b. $\mathbf{LOC}^k$, as a deterministic automata

Fig. 8. Transition $\sigma_3$ from $M$ to $x_5$, in $\mathbf{LOC}^k$

(b) $\sigma_4$ is a transition from $x_3$ to another state except $x_5$ (e.g. $x_i$): Since $\mathbf{LOC}^k$ is a deterministic automata, $x_i$ and $x_5$ must be control consistent states and $x_i, x_5$ become a new state $N$, constructed in $\mathbf{LOC}^k$. Two cases can be considered in Fig. 9.

(b.1) $s_3\sigma_4\sigma_5 \notin L(\mathbf{G})$

Assume $s_3\sigma_4\sigma_5 \notin L(\mathbf{G})$. Following the proof in Lemma 1, we see that, $\sigma_3$ is self-looped at all states in $\mathbf{LOC}^k$.

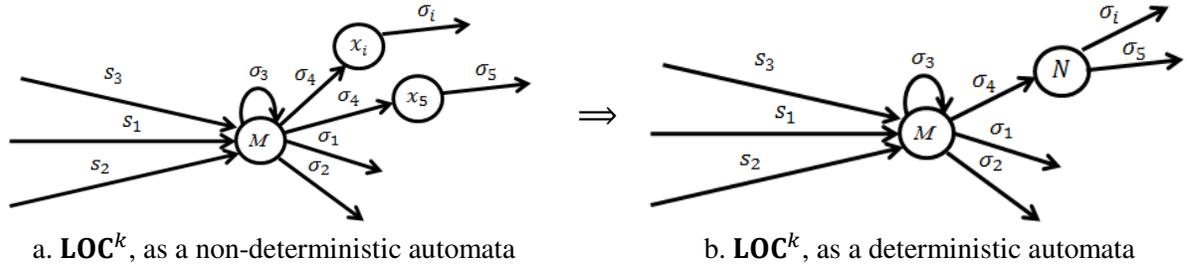

a. $\mathbf{LOC}^k$, as a non-deterministic automata

b. $\mathbf{LOC}^k$, as a deterministic automata

Fig. 9. Transition $\sigma_4$ from $M$ to $x_5$ and $x_i$ in $\mathbf{LOC}^k$

(b.2) $s_3\sigma_4\sigma_5 \in L(\mathbf{G})$

By substituting $\sigma_4$ by $\sigma_4\sigma_5$, (23) and (24) can be rewritten. Thus, the claim is proved by an iterative procedure.

We summarize the above argument in the following statement:

If an event $\sigma_3$ which is not disabled in $\mathbf{LOC}^k$, is self-looped at one state, e.g. $M$, then some other events may be self-looped at $M$, until we can find some enabled events at a control consistent state with $M$, such that they are not defined at corresponding state in $\mathbf{G}$. Thus, there is a natural projection $P: \Sigma^* \to \Sigma_0^*, \Sigma_0 = \Sigma_i \cup \{\sigma_3\}$, such that $P(s) = P(s')$, $s\sigma \in L(\mathbf{LOC}^k) \cap L(\mathbf{G}), s' \in L(\mathbf{LOC}^k), s'\sigma \in L(\mathbf{G}) \Rightarrow s'\sigma \in L(\mathbf{LOC}^k)$, and $s'\sigma \notin L(\mathbf{G})$.

□

*Remark 1:* The reverse of Theorem 1 is not true. It means that, the event reduction may occur in a local controller, but the state reduction may not.

We will show in example 4.2 that, the number of events used in each local controller is less than the event cardinality of the reduced supervisor (according to Corollary 1), but the state size of each local controller is same as the state size of the reduced supervisor. However, we prefer a set of local controllers, with fewer states than the reduced supervisor; each local controller needs fewer events comparing to the number of events used by the reduced supervisor in order to make consistent decisions. This gives us new criteria for evaluating the localizability of a supervisor. We can relax the definition of localizability according to event reduction criteria, i.e. a monolithic supervisor is localizable, if each local controller has fewer events comparing to the event cardinality of the reduced state monolithic supervisor.

*Corollary 2:* Comparing to a decomposable supervisor, we may find some local controllers corresponding to each component of the plant, such that each local controller has fewer events comparing to the reduced state original supervisor.

## 4. Examples

In this section, we consider examples in order to verify the extended theory in the paper.

### 4.1. Non-localizable case- Supervisory control of a shared common resource

The hard case in localization is where the component agents are coupled so tightly that each one has to be globally aware [11]. In Fig. 10 two agents $\mathbf{G}^i, i = 1,2$ share a common resource that is not allowed to be occupied simultaneously. It is easy to see that **SUP** is the recognizer of a monolithic supervisor which enforces the mutual exclusion specification. Then by applying the **localize** procedure in TCT software [19], we construct $\mathbf{LOC}^i, i = 1,2$ corresponding to each agent $\mathbf{G}^i$. We can see that each local controller is the same as **SUP** (Fig. 11). Thus, not only the state size is not reduced but also the event cardinality is not reduced. In this case, we say the monolithic supervisor is not localizable.

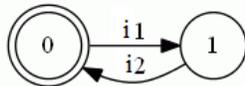

Fig. 10. DES model of agents, $\mathbf{G}^i, i = 1,2$

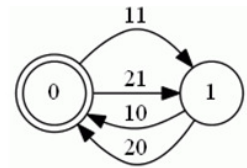

Fig. 11. The monolithic supervisor **SUP** and local controllers $\mathbf{LOC}^i, i = 1,2$

### 4.2. Localizable case in the term of event reduction criteria- Supervisory control of a guide way

Consider a guide way with two stations A and B, which are connected by a single one-way track from A to B on a guide way, as shown in Fig. 12. The track consists of 4 sections, with stoplights (*) and detectors (!) installed at various section junctions [18]. Two vehicles $V_1, V_2$ use the guide way simultaneously. $V_i$, $i = 1, 2$ may be in state 0 (at A), state $j$ (while travelling in section $j = 1, ...., 4$), or state 5 (at B). The generator of $V_i, i = 1,2$ are shown in Fig. 13.

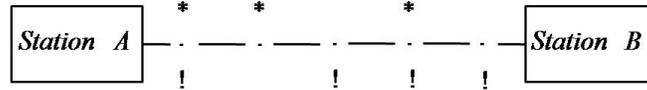

Fig. 12. Schematic of a guide way

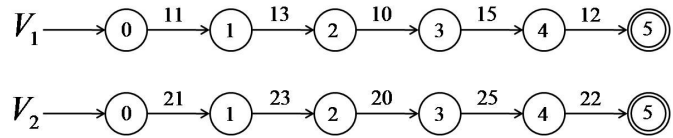

Fig. 13. DES model of each vehicle

The plant to be controlled is $G = \text{sync}(V_1, V_2)$. In order to prevent collision, control of the stoplights must ensure that $V_1$ and $V_2$ never travel on the same section of track simultaneously. Namely $V_i$, $i = 1,2$ are mutual exclusion of the state pairs $(i, i), i = 1,..,4$. The supremal relative observable supervisor and its corresponding reduced supervisor are shown in Figs. 14 and 15, respectively, where $P: \Sigma^* \to \Sigma_0^*$, $\Sigma_0 = \Sigma - \{13,23\}$.

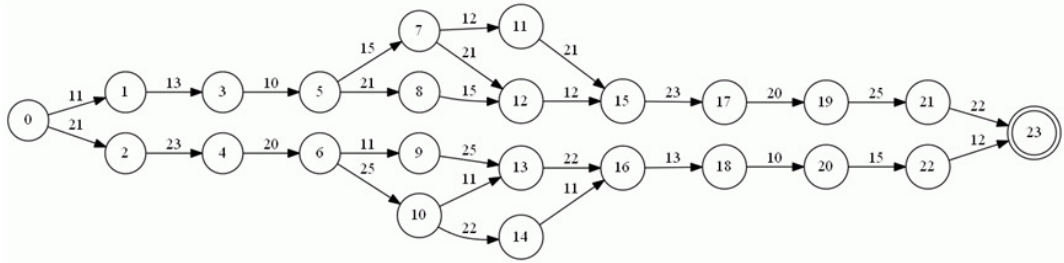

Fig. 14. The relative observable supervisor of guide way ($K_s$)

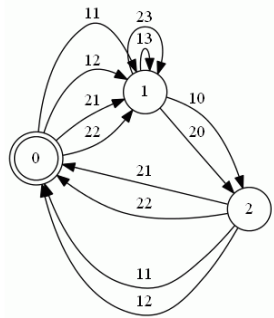

Fig. 15. The reduced supervisor of guide way ($K_r$)

Local controllers which are constructed by **localize** procedure are shown in Fig. 16. Since the number of states of each local controller is equal to the state cardinality of $\mathbf{K_r}$, the relative observable supervisor $K_s$ is not localizable, with state reduction criteria. But, we see the reduced supervisor employs some events which are not used by at least one local controller. For example, the reduced supervisor employs events 10, 12, 23 in decision making, but the local controller $\mathbf{V_1}$ do not use them. Also, the local controller $\mathbf{V_2}$ does not use events 13, 20, 22 in order to make decisions. Thus, we can say $\mathbf{K_s}$ is localizable in the term of event reduction criteria. Moreover, we can see that the monolithic supervisor $L_m(\mathbf{K_s})$ is decomposable w.r.t. $\mathbf{G}, P_1, P_2, \Sigma_1, \Sigma_2$, where $P_1: \Sigma^* \to \Sigma_1^*$, $\Sigma_1 = \Sigma - \{10,12,23\}$ and $P_2: \Sigma^* \to \Sigma_2^*$, $\Sigma_2 = \Sigma - \{13,20,22\}$.

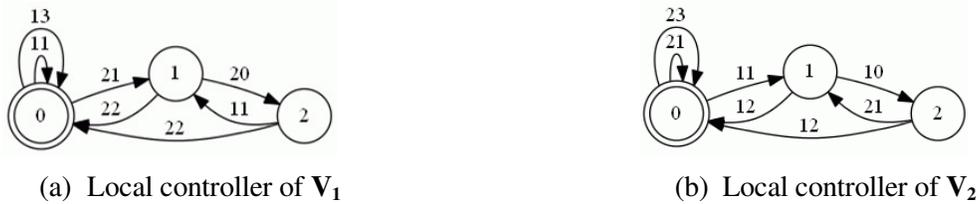

(a) Local controller of $\mathbf{V_1}$     (b) Local controller of $\mathbf{V_2}$

Fig. 16. Local controllers of each component of guide way

In the next example, we consider a case which the monolithic supervisor is localizable in the terms of state reduction and event reduction criteria. Moreover we will see that the number of events is reduced in each local controller, but the original supervisor is not decomposable.

### 4.3. Localizable case in the term of state reduction criteria- Supervisory control of transfer line

Industrial transfer line consisting of two machines $M_1$, $M_2$ and a test unit TU, linked by buffers $B_1$ and $B_2$, is shown in Fig. 17. If a work piece is accepted by TU, it is released from the system; if rejected, it is returned to $B_1$ for reprocessing by $M_2$. The specification is based on protecting the $B_1$ and $B_2$ against underflow and overflow [18]. All events involved in the DES model are $\Sigma = \{1,2,3,4,5,6,8\}$, where controllable events are odd-numbered. After the synthesis of the supremal controllable supervisor, we obtain the reduced supervisor by supervisor reduction procedure in TCT software, shown in Fig. 18. Moreover, we construct local controllers for each component $M_1$, $M_2$ and TU, as shown in Figs. 19-21.

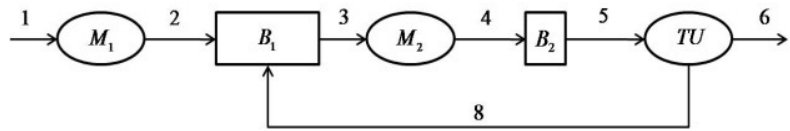

Fig.17. Transfer Line

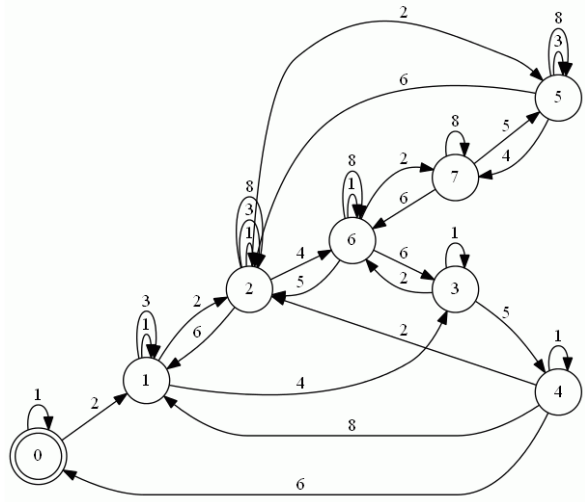

Fig. 18. Reduced supervisor for transfer line

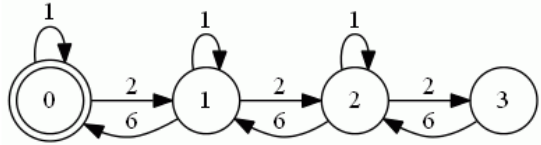

Fig. 19. Local controller $M_1$

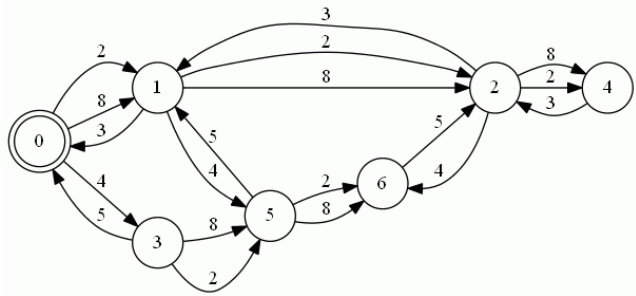

Fig. 20. Local controller $M_2$

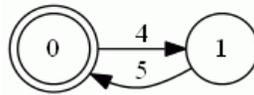

Fig. 21. Local controller TU

Obviously, we see that each local controller has lees number of states and less number of events, comparing to the reduced supervisor. In this example, the local controller $M_1$ does not use events 3, 4, 5, 8. Also, events 1, 6 are not used by the local controller $M_2$ and events 1, 2, 3, 6, 8 are not used by the local controller TU for decision making. Thus the synthesized supervisor for industrial transfer line is localizable in terms of state reduction and event reduction criteria. Whereas, we can see that the monolithic supervisor is not decomposable w.r.t. $\mathbf{G}, P_1, P_2, P_3, \Sigma_1, \Sigma_2, \Sigma_3,$ where $P_1: \Sigma^* \to \Sigma_1^*$, $\Sigma_1 = \Sigma - \{3,4,5,8\}$, $P_2: \Sigma^* \to \Sigma_2^*$, $\Sigma_2 = \Sigma - \{1,6\}$, and $P_3: \Sigma^* \to \Sigma_3^*$, $\Sigma_3 = \Sigma - \{1,2,3,6,8\}$.

## 5. Conclusion

This paper addresses the new property of event reduction in the supervisor localization procedure. We proved that if each local controller has less number of states comparing to the reduced supervisor, then such local controller has fewer events than the reduced state monolithic supervisor. In fact, localization procedure provides fewer states, for easier implementation of each local controller on industrial systems, and provides fewer events, in order to reduce the communication traffic between local controllers. The least communicated events between local controllers are important for efficiency of a distributed supervisory control. Thus, we can use the event reduction criteria for checking localizability of a supervisor. Localizability is satisfied by state reduction in each local controller, in the sense of [11], whereas event reduction satisfied the reduced communication traffic between local controllers, in the sense of this paper. By event reduction property, we can investigate and compare the localization and decomposition of a supervisor, in future work.

### References

[1] W. M. Wonham and P. J. Ramadge, "Modular supervisory control of discrete-event systems," Math. Control Signal Syst., vol.1, no.1, pp.13-30, 1988.
[2] R. J. M. Theunissen, M. Petreczky, R. R. H. Schiffelers, D. A. van Beek, and J. E. Rooda, "Application of Supervisory Control Synthesis to a Patient Support Table of a Magnetic Resonance Imaging Scanner", IEEE Trans. Autom. Sci. Eng., vol.11, no.1, pp.20-32, 2014.
[3] A. Afzalian, A. Saadatpoor and W. M. Wonham, "Systematic supervisory control solutions for under-load tap-changing transformers", Control Engineering Practice, vol.16, no.9, pp.1035-1054, 2008.
[4] Y. Willner and M. Heymann, "Supervisory control of concurrent discrete-event systems", International Journal of Control, vol.54, no.5, pp.1143-1169, 1991.
[5] J. Komenda, and J. H. Van Schuppen, "Modular Control of Discrete-Event Systems with Coalgebra", IEEE Trans. Autom. Control, vol.53, no.2, pp.447-460, 2008.
[6] F. Lin and W. M. Wonham, "Decentralized control and coordination of discrete-event systems with partial observation", IEEE Trans. Autom. Control, vol.35, no.12, pp.1330-1337, 1990.
[7] L. Feng and W. M. Wonham, "Supervisory Control Architecture for Discrete-Event Systems", IEEE Trans. Autom. Control, vol.53, no.6, pp.1449-1461, 2008.
[8] H. Marchand and S. Pinchinat, "Supervisory control problem using symbolic bisimulation techniques", in Proc. of Amer. Control Conf., Chicago, IL, June. 2000, pp. 4067–4071.
[9] A. Vahidi, M. Fabian, B. Lennartson, "Efficient supervisory synthesis of large systems", Control Engineering Practice, vol.14 (10), pp.1157-1167, 2006.
[10] R. Su and W. M. Wonham, "Supervisor reduction for discrete-event systems", Discrete-event Dyn. Syst., vol.14, no.1, pp.31–53, 2004.
[11] K. Cai and W. M. Wonham, " Supervisor Localization: A Top-Down Approach to Distributed Control of Discrete-Event Systems", IEEE Trans. Autom. Control, vol.55, no.3, pp.605-618, 2010.
[12] K. Cai and W. M. Wonham, "New results on supervisor localization, with case studies", Discrete-event Dyn. Syst., vol.25, no.1, pp.203-226, 2015.

[13] K. Rudie and W. M. Wonham, "Think globally, act locally: decentralized supervisory control", IEEE Trans. Autom. Control, vol.37, no.11, pp.1692-1708, 1992.
[14] F. Lin, W. M. Wonham, On observability of discrete-event systems, Information Sciences, 44(1988), 173-198.
[15] R. Cieslak, C. Desclaux, A. S. Fawaz, P. Varaiya, Supervisory control of discrete-event processes with partial observations, IEEE Trans. Autom. Control, 33(1988), 249-260.
[16] K. Cai, R. Zhang, W.M. Wonham, Relative Observability of Discrete-Event Systems and its Supremal Sublanguages, IEEE Trans. Autom. Control, 60(2015), 659- 670.
[17] C.G. Cassandras and S. Lafortune, "Introduction to Discrete-event Systems", 2nd edn. Springer, NewYork, 2008.
[18] W. M. Wonham, "Supervisory control of discrete-event systems", Lecture Notes University of Toronto, 2014, http://www.control.utoronto.ca/DES.
[19] W. M. Wonham, Control Design Software: TCT. Developed by Systems Control Group, University of Toronto, Canada, Updated 1st July 2014, http://www.control.utoronto.ca/cgi-bin/dlxptct.cgi.